\newcommand{\CLASSINPUTtoptextmargin}{0.73in}
\newcommand{\CLASSINPUTbottomtextmargin}{1.01in}
\documentclass[conference]{IEEEtran}
\IEEEoverridecommandlockouts
\usepackage{cite}
\usepackage{dirtytalk}
\usepackage{algorithmic}
\usepackage{graphicx}
\usepackage{amsmath}
\usepackage{amssymb}
\usepackage{balance}
\usepackage{textcomp}
\usepackage{multirow}
\usepackage{xcolor}
\usepackage{amsfonts}

\usepackage{siunitx}
\usepackage{gensymb}
\usepackage[english]{babel} 
\usepackage{blindtext}
\newcommand{\RNum}[1]{\uppercase\expandafter{\romannumeral #1\relax}}
\newcommand\mc{\mathcal}
\DeclareMathAlphabet{\pazocal}{OMS}{zplm}{m}{n}

\usepackage{mathtools}
\usepackage{url}

\def\BibTeX{{\rm B\kern-.05em{\sc i\kern-.025em b}\kern-.08em
    T\kern-.1667em\lower.7ex\hbox{E}\kern-.125emX}}
\begin{document}

\title{Machine Unlearning for Uplink Interference Cancellation
}

\author{\IEEEauthorblockN{Eray Guven\normalsize, Gunes Karabulut Kurt \normalsize} 
\IEEEauthorblockA{{\normalsize Poly-Grames Research Center, Department of Electrical Engineering,  Polytechnique Montr\'eal, Montr\'eal, Canada}
}
E-mail: guven.eray@polymtl.ca,
gunes.kurt@polymtl.ca
}

\maketitle

\begin{abstract}
Machine unlearning (MUL) is introduced as a means to achieve interference cancellation within artificial intelligence (AI)-enabled wireless systems. It is observed that interference cancellation with MUL demonstrates $30\%$ improvement in a  classification task accuracy in the presence of a corrupted AI model. Accordingly, the necessity for instantaneous channel state information for existing interference source is eliminated and a corrupted latent space with interference noise is cleansed with MUL algorithm, achieving this without the necessity for either retraining or dataset cleansing. A Membership Inference Attack (MIA) served as a benchmark for assessing the efficacy of MUL in mitigating interference within a neural network model. The advantage of the MUL algorithm was determined by evaluating both the probability of interference and the quantity of samples requiring retraining. In a simple signal-to-noise ratio classification task, the comprehensive improvement across various test cases in terms of accuracy demonstrates that MUL exhibits extensive capabilities and limitations, particularly in native AI applications. 
\end{abstract}
\section{Introduction} 
Fast model generation, adaptation, and recovery is essential in native artificial intelligence (AI) based ultra-reliable communication. In native AI based wireless communication systems, real-time data collection and model update policies becomes the part of the networks. Hence, any flaws or corruption in a data-driven model are contained, rendering it essentially useless. For this reason, re-utilization of AI models for valuable but scarce information in a corrupted model is a focus of area in machine learning applications \cite{ji2018model}.

Inter-user interference and adjacent-channel interference are two major reason of data and model corruption in wireless communication. In wireless uplink transmission, multi-user detection (MUD) and interference cancellation is required for signal recovery. Particularly during dataset collection, interference serves not only as a source of noise but also compromises the privacy of nonparticipating users. This results with corrupted (or poisoned) dataset and as well as the model generated with the corresponding dataset. 

In AI-powered wireless communication, should a data-driven model (e.g. a neural network) become corrupted, creating a new model is required to tackle the mentioned issues concerning privacy and performance. This necessities either a newly collected dataset or a cleaning algorithm in the dataset. In contrast to all this effort, machine unlearning (MUL) \cite{bourtoule2021machine} enables to forget a sample space correspondence in a latent space, which will be referred as model cleansing throughout the article. For interference exposed wireless communication models, MUL can be utilized for model cleansing on a corrupted model without the need for real-time source information or identification, effectively canceling the interference. As a result, not only a secure and private but more accurate AI model construction is possible with MUL with less cost. The most importantly, a regeneration or a pre-processing on the dataset is not required. As far as authors' knowledge, there is no prior study regarding a MUL based interference cancellation in the literature. Model cleansing in uplink communication can be necessitated by several factors:
\begin{enumerate}
    \item \textbf{User Privacy Policy}: Upholding user privacy policies mandates the cleansing of models to remove any sensitive information inadvertently captured during training.

    \item \textbf{Untrusted Sources}: Models may require cleansing when trained on data from untrusted sources to mitigate the risk of incorporating malicious or erroneous information.

    \item \textbf{Avoiding Lengthy Retraining and Dataset Acquisition}: Cleansing models eliminates the need for extensive retraining and acquiring new datasets, streamlining the optimization process and reducing resource consumption. 
\end{enumerate}
Several use cases of MUL and its verifiability are detailed in \cite{xu2024machine}, presented within the explainable AI framework.  


While conventional interference cancellation methods typically rely on precise identification of interference parameters such as the amount of interference, time interval, and optionally, channel state information (CSI) for accurate cancellation \cite{costaa}, MUL necessitates only soft detection of interference time intervals, a task achievable with any MUD methods \cite{verdu1998multiuser}. Despite being a new concept in wireless communication, machine unlearning already has some practical uses in this field. In \cite{zhang2022poison}, MUL is used to strengthen models against backdoor attacks in mmWave beam prediction applications. Authors in 
\cite{wu2022federated} provides user privacy in distributed networks with MUL integrated federated learning. Furthermore, MUL is proposed for the security of neural networks against backdoor injection attack in \cite{liu2022backdoor}. 

Deep learning aided interference mitigation techniques are quite popular and summarized in \cite{oyedare2022interference}. However, the current methods are solely based on either learning the adaptation on interference existing environment, or dataset based counter-poisoning algorithms \cite{wan2024data}. This study utilizes the machine unlearning for model cleansing for the inter user interference on uplink transmission. By doing that, not only the privacy of user(s) out of interest is preserved, but also the accuracy of the model is increased for the given task due to the mitigated interference. Accordingly, a membership inference attack (MIA) based performance metric for the cross validation in presented. MIA poses a significant threat to user privacy as it can potentially reveal whether an individual's data was used to train a machine learning model \cite{shokri2017membership}.

Data poisoning for adversarial attacks is another major challenge on ML applications. Even though there are many methods for data cleansing such as utilizing machine learning, model cleansing has many advantages over the dataset cleaning as it does not require the dataset itself. An interference source during real time dataset gathering is a major problem in AI enabled communication networks. Inter user interference, just one type of this interference's, is examined in this study. As a simple use case, the latent space representation of randomly adjointed user interfered captured in-phase and quadrature (IQ) to signal-to-interference-plus-noise ratio (SINR) information will be performed. This objective also will serve as a benchmark for machine unlearning. Therefore, as long as the interfered model undergoes cleansing, the expected outcome is a reduction in the difference between posterior distributions. MUL method tunes a signal-based model, regardless the purpose of the model, without drifting away from the inherit characteristics of the model. In this study, we utilize MUL on a model which exposed to different severity level of interference. 
\begin{figure}[] 
    \centering
    \includegraphics[width=0.40\textwidth]{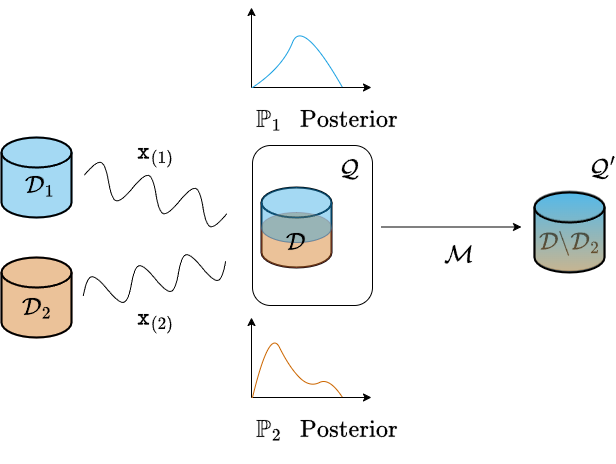}
    \vspace{-2mm}
    \caption{The goal of the MUL to find a $\mc{Q}'$ to reach minimum loss in the original task. A MIA questions the ownership of input sample of $\mathtt{x}$ by using $\mathbb{P}_1$ and $\mathbb{P}_2$.}
    \vspace{-6mm}
    \label{muldiagram}
\end{figure}
The main contribution of this study is the MUL introduction on model cleansing within interference exposed AI models. This study presents MUL as a groundbreaking solution for interference cancellation, offering significant gains in both resource and time efficiency. A corrupted dataset generated model is cleansed and a cross-validation is performed with the MIA. The validity of MUL is tested with a signal-to-noise ratio (SNR) classification example under several interference conditions. We found that posterior distribution of sample loss in interfered cancelled model approaches to only retain-set learned model's sample loss posterior distribution. with sufficient tuning iterations.

\section{Preliminaries and Problem Construction}
\label{sec2}
Consider an uplink communication scenario with single antenna $N$ users and single antenna base station (BS). The modulated symbol $\textbf{x}_1 \in \mathbb{C}$ with normalized unit power $\mathbb{E}[|\textbf{x}_1|^2]=1$ is being transmitted to the BS from a constant user source $U_1$ within a time interval $t_0\leq t \leq t_1$. During this period, only active user is $U_1$ and the received signal $\textbf{y} \in \mathbb{C}$ at BS is following
\begin{equation}
    \textbf{y}_1 = \textbf{h}_1 \textbf{x}_1 + \textbf{n} 
\end{equation}
where the independent and uncorrelated frequency-selective fading channel is $\textbf{h}_1 \in \mathbb{C}$ and $\textbf{n} \sim \mathbb{CN}(0,\sigma^2)$ is i.i.d additive complex white Gaussian noise. In this case, SNR \textcolor{black}{of $U_1$} with unit power is following
\begin{equation}
    \textrm{SNR}_{(1)} = \frac{|\textbf{h}_1|^2}{\sigma^2},
\end{equation}
There exist a second time interval $t_1 < t \leq t_2$ where $U_{j\neq 1}$ is also active and interferes the BS during the dataset construction, resulting with the $\textbf{y}_N$ in BS as follows 
\begin{equation}
    \textbf{y}_N = \sum_{n=1}^N \textbf{h}_n \textbf{x}_n + \textbf{n}_N 
\end{equation}
Thereby, the signal-to-noise and interference ratio (SINR) on the second time interval is following
\begin{equation}
    \textrm{SINR}_{(N)} = \frac{|\textbf{h}_1|^2}{\displaystyle{\sum_{k\neq 1}{|\textbf{h}_k|^2}} + \sigma^2_N},
\end{equation}
With the assumption of $|\textbf{h}_1|^2>|\textbf{h}_k|^2$, a successive interference cancellation (SIC) algorithm \cite{sen2010successive} can be applied to capture $U_1$ throughout $t_0$ to $t_2$. However, this requires the knowledge of amount of interference users and channel state information (CSI) on both $U_1$ and $U_{j\neq 1}$. Considering the extensive time interval, it is a costly application under fast fading channels. On the other hand, a MUL algorithm is capable to cancel the interference with only a non-interfered retain set. 

\subsection{Model and Dataset Construction}
A supervised SNR classification application is under investigation as an example for MUL interference cancellation. The dataset $\mathcal{D}_U = \left\{\mathtt{x}^{(i)}, \gamma^{(i)} \, \middle| \, i \in \left\{1, 2, \ldots, T\right\}\right\}
$ where $\gamma^{(i)} \in \mathbb{R}^1$ is the SINR of the $i$\textsuperscript{th} block and the feature map of the received modulated information is denoted as $\mathtt{x} \in \mathbb{R}^2$. 2-dimensional IQ feature maps captured with follows
\begin{equation}
    \mathtt{x}_s^{2 \times T} =
     \big[\mathrm{Re}\{\textbf{y}_n\}~;~\mathrm{Im}\{\textbf{y}_n\}\big].
\end{equation}
Each $A^2$ map batch is obtained with $\mathtt{x}_s^{2 \times T} \rightarrow \mathtt{x}^{(2T/A^2) \times A \times A}$ which corresponds to $i\textsuperscript{th}$ block transmission. Structure of $\mc{D} = \{\mc{D}_1~;~ \mc{D}_2\}$ is follows

\begin{equation} \mathcal{X}= 
    \begin{cases}
    \mathtt{x}_{(1)} \subset \mathcal{D}_{1}, & \text{Retain Set},\\
    \mathtt{x}_{(N)} \subset \mathcal{D}_2 , & \text{Forget Set} 
    \end{cases}
\end{equation}
\begin{equation}  
    \mathcal{Y}   = 
    \begin{cases}
    \textrm{SNR}_{(1)} \subset \mathcal{D}_{1}, & \text{Retain Set},\\
    \textrm{SINR}_{(N)} \subset \mathcal{D}_2 , & \text{Forget Set} \\
    \end{cases}
\end{equation}
An evaluation dataset $\mc{D}_{\textrm{test}}$ follows the same procedure is for the evaluation. In contrast to $\mc{D}$, several $\mc{D}_{\textrm{test}}$ exist with different interference severity to evaluate the validity of the MUL. 

\section{Machine Unlearning}
We define a neural network based predictive model as follows: a data driven generated latent space $\mathcal{F}$ aims to map the samples $\mathtt{x}$ in sample space $\mathcal{X}$ to the corresponding output $\gamma$ to output space $\mathcal{Y}$ as in $\mathcal{F}:\mathcal{X}\rightarrow \mathcal{Y}$.   
For the corresponding latent space, a training subset $\mc{D}_{\text{train}}$ and test subset $\mc{D}_{\text{test}}$ forms each $\mc{D}$. The sample space for this study is defined as transmitted information from active $N$ user(s) to a BS. In a conventional uplink scenario, the collected dataset is denoted as $\mc{D}_U$ and $\mc{D}_U'$ is defined as the interference cancelled dataset, $\mathcal{D}$ construction with two users on uplink is defined as following
\begin{align}
    \mathcal{D}_{U} &:= U_1 \oplus U_2,\\
    \mc{D}_U' &:= U_1 \oplus U_2 \ominus \tilde{U}_2,
\end{align}
where $\textbf{y}_1 \in U_1$, $\textbf{y}_2 \in U_2$ and $\tilde{U}_2$ is the estimated $U_2$ in BS. The latent space is generated by an AI model $\mathcal{Q}$. The MUL process $\mathcal{M}$ aims $\mathcal{Q}\xrightarrow{\mathcal{M}}\mathcal{Q}'$ such that $\mathcal{Q}'$ is the cleansed $\mc{Q}$ with minimized interference exposure. An ideal $\mathcal{M}$, the result model $\mathcal{Q}'$ is not equal to $\mathcal{Q}$. In order to ignore the estimation errors on $U_2$ detection, $\mc{D}_U' = U_1$ is established, which also considered as \textit{retain set} whereas $\mc{D}_2 = U_2$ is defined as \textit{forget set}. 



\subsection{Evaluation}
A simple way to evaluate the performance of $\mathcal{M}$ is to utilize $\mathcal{D} \backslash d_u$ to regenerate $\mathcal{Q}^\circ$ such that $d_U$ has no impact in it. Accordingly, $\mathcal{Q}^\circ$ is denoted as re-learned retain (RR) model. In this case, $\mathcal{Q}^\circ$ can be interpreted as the ground truth model for $\mathcal{Q}'$. Therefore, $\mc{M}$ results with $\mathcal{Q}' = \mathcal{Q}^\circ$ in an ideal unlearning process and it is desired that $\mc{Q}'$ perform as well as $\mc{Q}^\circ$ for the same sample space $\mc{X}$. Thence, $\mc{D}_{\text{train}}$ and $\mc{D}_{\text{test}}$ can be used in both $\mc{Q}$ and $\mc{Q}'$ for a fair comparison. For the interference cancellation, $d_U = \mc{D}_2$. Moreover, previously established $\mc{D}_U' = U_1$ allows to conserve $ \mc{D}_U' = \mc{D} \backslash d_U$. In one point of view, the interference source of $U_2$ is considered as the undesired forget set.  

\subsection{Unlearning with fine-tuning}
Unlearning (or forgetting) the interference on $\mc{Q}$ can be done in several ways.
As a  simple MUL method, unlearning by fine-tuning ($\mc{M}_F$) the instance model $\mc{Q}$ with $\mc{D}_{1}$ can be performed to generate $\mc{Q}'$. Therefore, one interpretation of the unlearning $\mc{D}_{2}$ within $\mc{Q}$ is the fine-tuning the model parameters with the retain set of $\mc{D}_{1}$. 

Other than inter-user-interference, an unknown source can also interfere $\mc{D}$, either unintentionally or deliberately as an adversary. In this case, the corresponding forget set is unknown. The advantage of $\mc{M}_F$ is that $d_U$ is not required during the unlearning. In this way, non-identified interference sources and unknown noises are cancelled as well. It's noted that channel gain conditions in SIC are also eliminated. The most importantly, $\mc{M}_F$ is not restricted by the severity of interference but the duration of it. 

\subsection{SNR Classification}
In the SNR classification task, an interference with constant power spectrum sourced from $U_2$ interferes $\mc{D}_U$. Uniformly distributed random bits are QPSK modulated and passed through a Rayleigh fading channel with zero mean and evenly distributed phase. For the simplicity, large scale fading does not exist for any user and non-orthogonality of users are provided with narrowband transmission power difference, which remains same during the $\mc{D}$ collection. Each user subject to the same noise spectral density and assuming that received signal is fully synchronized and channel state information of $U_1$ is perfectly known by BS, received signal is recovered with zero forcing equalizer. Any interference cancellation during data collection is not employed and $\mc{Q}$ is generated with $\mc{D}_U$ as illustrated in Figure \ref{muldiagram}.

\subsection{Adversary Model}
An adversary attack reveals if $\mc{Q}'$ carries $d_U$ information and answers the following questions:
\begin{itemize}
    \item Is $\mc{Q}'$ more or less secure than the original model $\mc{Q}$?
    \item Is $d_U$ is completely removed from the $\mc{Q}$?
    \item Does $\mc{M}$ expose the $\mc{D}_2$ privacy?
    \item What happens if $d_U >> \mc{D}_2$?  
\end{itemize}
MIA is a simple benchmark to evaluate the quality of unlearning, and therefore $\mc{M}$. Using the posterior distribution of the sample loss, MIA can evaluate the ownership probability of the model with a comparison of target and forget losses. Accordingly, each sample loss in the output of $\mc{Q}$ and $\mc{Q}'$ by using $\mc{D}_{\text{train}}$ and $\mc{D}_{\text{test}}$.   
Interference cancellation with MUL for ensemble models is also possible. Similarly, each subdataset for a given task is $\mc{D}_U^{e,j}$. Now, we set ensembled datasets of $\mc{D}^e_U = \{\mc{D}_U^{e,1}, \mc{D}_U^{e,2},\cdots,\mc{D}_U^{e,S},\cdots,\mc{D}_U^{e,M}\}$ where the only interfered dataset is $\mc{D}_U^{e,S}$, and the generated dataset is $\mc{Q}_U^e$. Instead of retraining the $\mc{Q}_U^e$, one can unlearn $\mc{Q}_U^{e,S}$ to form a modified model $\mc{Q}'^{e}_U = \{\mc{Q}_U^{e,1}, \mc{Q}_U^{e,2}, \cdots, \mc{Q}'^{e,S}_U, \cdots, \mc{Q}_U^{e,M}\}$. 
\begin{figure*}[] 
    \centering \includegraphics[width=1\linewidth]{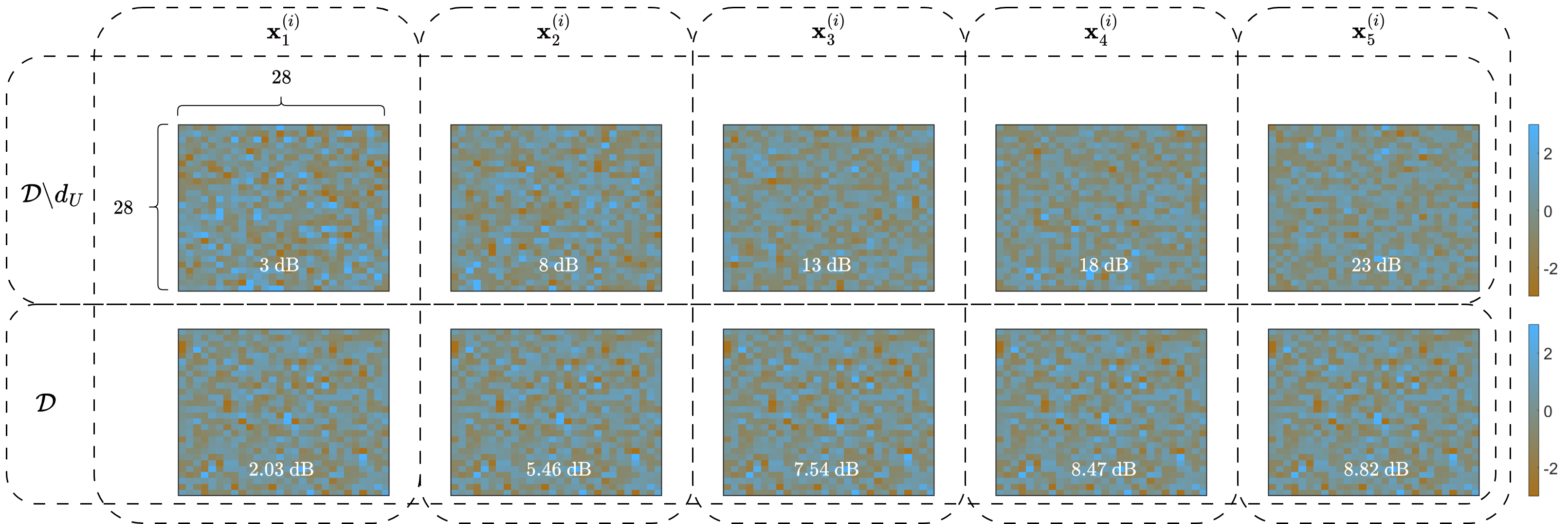}
    \vspace{-6mm}
    \caption{A sample of each class of IQ feature maps within $\mc{D}$ under interference and no interference. An unlearning model of $\mc{M}$ fine-tunes the model $\mc{Q}$ to cancel out the interference, without retrain the retain set $\mathcal{D} \backslash d_U$. Increasing noise and interference causes deterioration on the uniformity of symbol distrubution and energy per map.}
    \vspace{-5mm}
    \label{direc}
\end{figure*}

\subsection{Data Collection and Unlearning Cost}
In this section, the interfered uplink signal collection and the unlearning cost for this dataset can be found in terms of the number of samples needed to be unlearned. To continue with the $N$ user case scenario, the amount of MUL request is $N-1$. It is assumed $\mc{D}_2 << \mc{D}_1$ so that the unlearning process does not change the size of $\mc{D}$. $U_1$ always exists on each time frame and the probability of the $i$-th user is active with ${\mc{P}_{i+1}}$. Due to the superimposed operation on interference, the number of users does not affect the total sample number $K$.  Furthermore, the amount of sample needs to be training is $K - \sum_{i=1}^M \sum_{j=1}^N d_{U_j}^{(i)}$ for $M$ amount of model. Assuming the same amount of samples is collected within each uniform
$t_0$ time frame, corresponding to $\frac{K}{t_0}$ samples for each time frame. Therefore, the probability of $\mc{D}$ construction with $N \geq 2$ users is $\prod_{i=1}^{N-1} {{\mc{P}_{i+1}}}$ in a time frame. The event of at least one interference exist within a model is denoted as $E_I$ and it's probability is given as below
\begin{equation}
    \mc{P}_{E_I} = 1 - \prod^{N-1}_{i=1} \left(1-{\mc{P}_{i+1}}\right) 
\end{equation}
Assuming each time frame is independent from another, the probability of $m$ out of $M$ models having IUI by $N-1$ users can be shown with binomial distribution, where the probability mass function of $m = \bigcup^{N-1}_{i=1} U_{i+1}$ for $0 \leq m \leq M$ is given below
\begin{equation}
 \mathbb{P}_X(m)=\left(\begin{array}{c}
M \\
m
\end{array}\right) (\mc{P}_{E_I})^m (1-\mc{P}_{E_I})^{M-m}
\end{equation}

In a scenario where each user interference detected, $N-1$ unlearning request is required. For each user interference unlearning, first unlearning request, $(\frac{K}{t_0}-1)$ sample needs to be retrained. One the $i$-th request, it is $(\frac{K}{t_0} -i)$ samples needs to be retrained with $\frac{1}{t_0}$ probability in the same submodel and $(\frac{K}{t_0}-1)$ samples need to be retrained on any other submodels with $1 - \frac{1}{t_0}$ probability. On $U_3$, a submodel with only interfered by previous users (which is only $U_2$ in this case) are not subject to unlearning again. First unlearning request $(\frac{K-K'}{t_0-t'_0}-1)$ retraining where $K'$ is the cleansed submodel(s) sample number from the previous unlearning and $t'_0$ is the length of time corresponding cleansed submodel(s). Similarly, on the $i$-th request of $U_3$ interference, $(\frac{K-K'}{t_0}-i)$ will be retrained with $\frac{1}{t_0-t'_0}$ probability if the submodel is the same with first unlearning request on $U_3$ cancellation. Otherwise, $(\frac{K-K'}{t_0-t'_0}-1)$ samples needs to be retrained with $1-\frac{1}{t_0-t'_0}$ probability. This process continues until $m$ defected submodels are cleansed.
\begin{figure*}
    \centering  \includegraphics[width=1\textwidth]{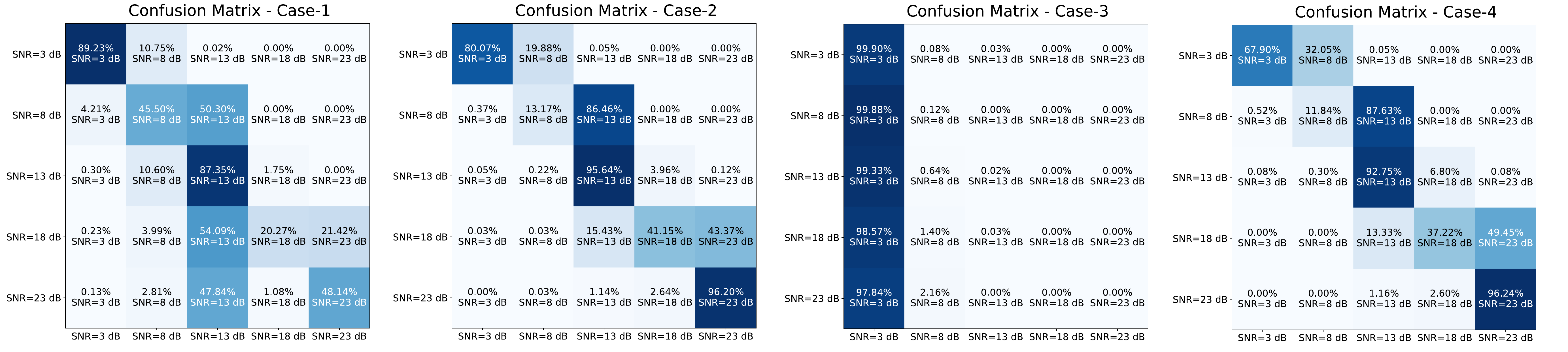}
            \vspace{-5mm}
    \caption{SNR classification in the original model $\mc{Q}$ in different severity of interference.}
    \label{conf}
    \vspace{-1mm}
\end{figure*}
\begin{figure*}
    \centering  \includegraphics[width=1\textwidth]{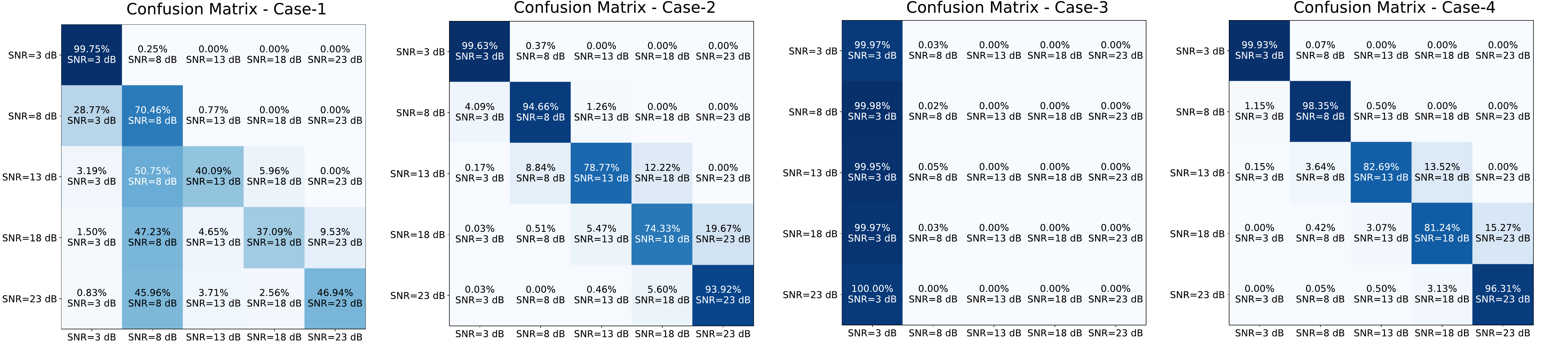}
            \vspace{-5mm}
    \caption{SNR classification in the unlearned model $\mc{Q}'$ in different severity of interference.}
    \label{conf2}
        \vspace{-2mm}
\end{figure*}
On the first user cancellation, the probability of $\frac{K}{t_0}-1-j$ sample in retraining in $i$-th request after the first user cancellation is following
\begin{equation}
 \left(\begin{array}{c}
i-1 \\
j
\end{array}\right) \Bigl(\frac{1}{t_0}\Bigl)^j \Bigl(1-\frac{1}{t_0}\Bigl)^{i-j-1}
\end{equation}
Likewise, after $x$ user interference cancellation, the probability of $\frac{K-K'}{t_0 - t'_0}-1-j$ is following for $0\leq j \leq i-1$
\begin{equation}
\left(\begin{array}{c}
i-1 \\
j
\end{array}\right) \Bigl(\frac{1}{t_0 - t'_0}\Bigl)^j \Bigl(1-\frac{1}{t_0 - t'_0}\Bigl)^{i-j-1}
\end{equation}
Therefore, the total amount of unlearned samples in each submodel (first sum) of the each request (second sum) is
\begin{equation}
    \mathbb{E}[A_1]=\sum_{i=1}^{K} \sum_{j=0}^{i-1} \left(\begin{array}{c}
i-1 \\
j
\end{array}\right) \Bigl(\frac{1}{t_0}\Bigl)^j \Bigl(1-\frac{1}{t_0}\Bigl)^{i-j-1} \Bigl(\frac{K}{t_0}-1-j\Bigl) 
\end{equation}
After $\nu$ user interference cancellation where $\nu \leq N-2$, the total amount of unlearned samples in each submodel (first sum) of the each request (second sum) of the each user (third sum) is the following
\begin{align}
    \mathbb{E}[A_\nu]=\sum_{n=0}^{\nu} &\sum_{i=1}^{K}   \sum_{j=0}^{i-1} \left(\begin{array}{c}
i-1 \\
j
\end{array}\right) \Bigl(\frac{1}{t_0-t^{(n)}_0}\Bigl)^j \nonumber \\
&\times \Bigl(1-\frac{1}{t_0-t^{(n)}_0}\Bigl)^{i-j-1} \Bigl(\frac{K-K^{(n)}}{t_0-t^{(n)}_0}-1-j\Bigl)  
\end{align}

\section{Numerical Results}
IQ feature map is taken as grey scale image with $1$ input channel in the first convolution layer, and it outputs $32$ channel by using $3\times3$ kernel. With forward pass data flow, first fully connected layer produces $128$ unit with $26\times26$ spatial size that following with $\mathrm{ReLU}$ activation function. Using the $128$ units from previous layer, the second fully connected layer obtains a $5$ dimensional logits corresponding to each class and a $\mathrm{Softmax}$ is applied the logits to obtain class probabilities. Learning is demonstrated with $0.001$ learning rate for $50$ iteration, optimizing with Adam optimizer. The classifier accuracy in each batch is the ratio of correct SNR estimations to the total batch number. The complete model can be found in the soruce code \footnote{ 
    \url{https://github.com/riguwen/MULforIC.git}}.


\label{sec3}


$\mc{D}$ samples are generated with repetitive simulations with the given system model in Section \ref{sec2}. An SNR classification task is solely rely on the pretrained $\mc{Q}$, which is a weak model. A model cleansing applied on $\mc{Q}$ to generate $\mc{Q}'$ is expected to perform not only higher accuracy, but also does not violate of right of privacy for $U_{j\neq 1}$. A MUL application is performed with fine tuning on a corrupted model $\mc{Q}$ with interference. Regarding this, a custom dataset called Signal Interfered IQ dataset (SIIQ) is generated as follows. Within the $t_0<t<t_1$ interval with $\tau_1$ duration, $T = 3125$ retain samples forms the non-interfered signal while the second interval $t_1<t<t_2$ with $\tau_2$ duration contains $T = 625$ forget samples which forms the interfered signal. The use of $\mc{Q}'$ can be evaluated with a number of different case interference severity. \\

\noindent \textbf{CASE-1- Weak Interference}: $N=2$ where $U_2$ with $-12$ dB SINR on the BS such that $\gamma_2 < \gamma_1$ during the dataset $\mc{D}_U$ collection In the weak interference case, $\tau_2 / \tau_1 = 5$. \\
\noindent \textbf{CASE-2- Strong Interference:} $N=2$ where $U_2$ with $-4$ dB SINR on the BS such that $\gamma_2 < \gamma_1$ during the dataset $\mc{D}_U$ collection. In this case, the interference detection is easier yet corruption is more severe. In the strong interference case, $\tau_2 / \tau_1 = 5$. \\
\noindent \textbf{CASE-3- Persisting Interference:} $N=3$ where $U_{2,3}$ with $-12$ dB SINR on the BS such that $\gamma_2 < \gamma_1$ during the dataset $\mc{D}_U$ collection. The duration of each interference is independent, which requires substantial weight optimization on $\mc{Q}$. In the persisting interference case, $\tau_2/\tau_1 = 1$. \\
\noindent \textbf{CASE-4- No Interference} Within this dataset, the interference does not exist. Each class of signal powers are uniformly distributed. Consequently, $\tau_2/\tau_1 = \infty$. 
\begin{table}[h]  
\vspace{-4mm}
\caption{Accuracy of original ($\mc{Q}$), RR ($\mc{Q}^\circ$) and unlearned ($\mc{Q}'$) model} 
\centering 
    \resizebox{0.7\linewidth}{!}{

\begin{tabular}{c c c c c} 
\hline\hline   
 Test Case & Subset  &$\mc{Q}$  &$\mathcal{Q}^\circ$ (RR) &$\mc{Q}'$ (MUL)
\\ [0.5ex]  
\hline   
  &train &66.56 &- &- \\[0ex]  
  &test &58.35 &59.11 &59.11  \\[-1ex]
\raisebox{1.5ex}{\textbf{Case-1}} & forget  
& -  &29.24 &29.24\\[0ex] & retain  
& - &88.43 &88.41 \\[1ex]  
  &train &66.56  &- &- \\[0ex]  
  &test &65.28   &88.49 &88.49\\[-1ex]
\raisebox{1.5ex}{\textbf{Case-2}} & forget  
& -  &29.24 &29.24 \\[0ex] & retain  
& -  &88.43 &88.41 \\[1ex]  
  &train &66.56  &- &- \\[0ex]  
  &test &19.62  &19.61 &19.61 \\[-1ex]
\raisebox{1.5ex}{\textbf{Case-3}} & forget  
& -  &29.24 &29.24\\[0ex] & retain  
& -   &88.43 &88.43\\[1ex] 
  &train &66.56  &- &- \\[0ex]  
  &test &61.05   &91.83 &91.83\\[-1ex]
\raisebox{1.5ex}{\textbf{Case-4}} & forget  
& - &29.24 &29.24\\[0ex] & retain  
& - &88.43 &88.43 \\[1ex] 
\hline \hline 
\end{tabular}  
}
    \vspace{-1mm}

\label{tabl1}
\end{table} 
Calculation of the cross-validation of the MIA for each case is based on computed loss $\mc{L}$. As long as $\mc{D}_{\text{train}}$ and $\mc{D}_{\text{test}}$ correlates with each other, the posterior distribution of the sample loss approximates in any model of $\mc{Q}_{0}$ such that $\mathbb{P}(\mc{L}(\mc{Q}_0)|\mc{D}_{\text{train}}) \approx \mathbb{P}(\mc{L}(\mc{Q}_0)|\mc{D}_{\text{test}})$. A logistic regression is chosen as an attack model, and each fold of cross validation on the attack model uses sample loss as input and member label ($0$ and $1$) as a target variable. As a consequence, a comparison between the test and forget sets provides a distinct indicator to infer whether a sample has been utilized in the model or not. 
\label{sec4}
\begin{figure}
\centering
\includegraphics[width=0.5\textwidth]{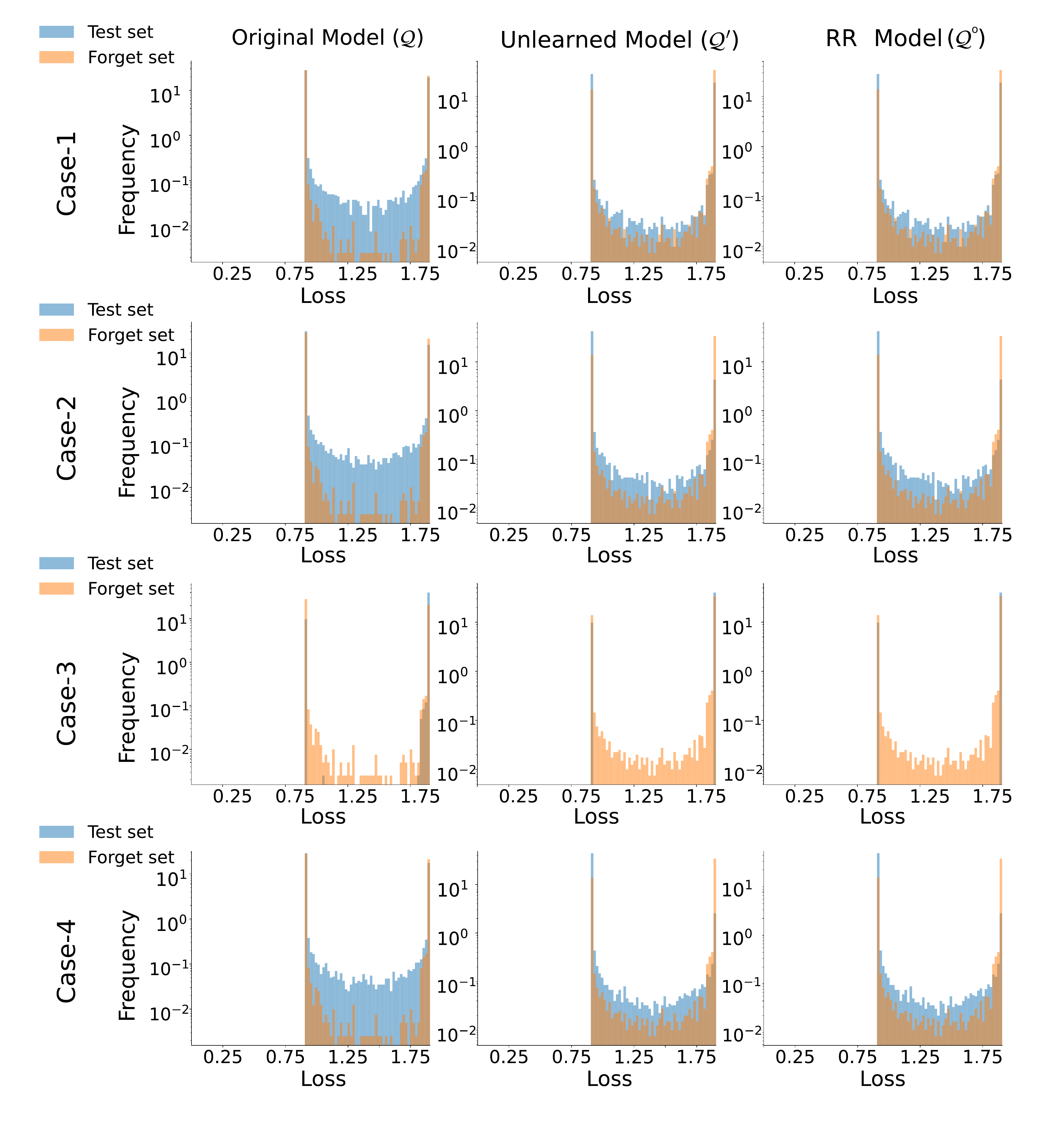}
\vspace{-7mm}
    \caption{Comparison of $\mathbb{P}_1$ and $\mathbb{P}_2$ in each model and cases. As the interference cancelled from $\mc{Q}$, forget loss increases and test loss decreases. Except for Case-3, the success of $\mathcal{M}_F$ is evident. }
    \vspace{-4mm}
    \label{fig:MULPost}
\end{figure}
According to accuracy results given in Table \ref{tabl1}, unlearning algorithm $\mc{M}_F$ successfully cancels the interference for Case-1 and Case-2 and Case-4. Regarding the Case-3, due to the high amount of interfered samples and scarce retain set, $\mc{M}$ unable to optimize the model $\mc{Q}$ further. Considering the accuracy is upper bounded by RR, $\mc{Q}'$ reaches to its maximum efficiency in Case-2 and Case-4. 

It is observed that, signal interfered forget subset indeed decreases the performance of the original model as $\mc{Q}$ reaches its maximum accuracy with $66.56\%$. In each case, once interference exposed weights are fine-tuned, the accuracy of the corrupted $\mc{Q}$ reaches up to $91.83\%$. It is also observed that there exist an margin of accuracy error, since the test accuracy and retain set accuracy on $\mc{Q}'$ has nearly $\sim 2\%$ differenced. 

\begin{table}[h]
    \vspace{-3mm}
    \centering
    \caption{MIA Scores}
    \vspace{-2mm}
    \resizebox{0.6\linewidth}{!}{
    \begin{tabular}{|c|c|c|c|}
        \hline
        \multirow{2}{*}{\textbf{Case}} & \multicolumn{3}{c|}{Scores} \\
        \cline{2-4}
         & $\mc{Q}$ & $\mc{Q}'$ & $\mc{Q}^\circ$ \\
        \hline
        1 & 0.50417 & 0.65204 & 0.65207 \\
        2 & 0.53779 & 0.79308 & 0.79323 \\
        3 & 0.68617 & 0.54888 & 0.54815 \\
        4 & 0.52125 & 0.81304 & 0.81555 \\
        \hline
    \end{tabular}
    }
         \vspace{-1mm}

    \label{tab:2}
\end{table}

Several conclusions can be drawn based on the MIA score. First conclusion is that each MIA score above $0.5$ implies that test and forget set samples follow different distributions. We easily verify this with the different SNR and SINR distributions on test and forget sets, due to the interference.

Figure \ref{fig:MULPost} illustrates that upon unlearning interference, the accuracy of classification for interfered samples decreases, consequently leading to an increase in the loss associated with the forget set. In contrast, it is expected to see that, test set loss to be decrease as the $\mc{Q}'$ presents a better representation for $U_1$ sample space. Therefore, it is expected to see that $\mathbb{P}_1$ and $\mathbb{P}_2$ converge in the interference unlearned cases. It is observed that, $\mc{M}_F$ satisfies this condition on all cases, except Case-3.  As a result, increasing MIA scores in Table \ref{tab:2} verifies the success of the $\mc{M}_F$ in Figure \ref{fig:MULPost}.




The confusion matrices for different cases are given in Figure \ref{conf}. Apart from Case-3, a noticeable enhancement in overall accuracy is evident.   As an example on Case-3, original model and MUL algorithm fails in SNR classification. It can be seen that every sample is misclassified as low-SNR signal due to the severe interference, even after $\mc{M}_F$. The Case-3 confusion matrix shows that there is such a limitation in $\mc{M}_F$ to be accurate and it is determined by $\mc{Q}$ success. 
\section{Conclusion}
In conclusion, MUL is presented for model-based interference cancellation. Models generated using  corrupted datasets by unknown interference sources were assessed for recovery, without modifying the dataset or re-generating the model. The investigation focuses on cleansing an IQ-based AI model without employing a real-time interference cancellation algorithm. The efficacy of this approach is validated through MIA cross validation and confusion matrices across various test cases. It is observed that any model affected by unknown sources of interference, can be further improved with MUL, with the condition of original model convergence. Therefore, the existing limitation is not caused by the interference severity, but $\mc{Q}$.

\label{sec5}

\bibliographystyle{IEEEtran}
\bibliography{reference}
\end{document}